\documentclass[draft,grl]{agutex}
\usepackage{lineno}
\usepackage{lscape}
\usepackage{graphicx}
\setkeys{Gin}{draft=false}
\authorrunninghead{VALENCIA-CARDONA ET AL.}
\titlerunninghead{Bullen's parameter and spin crossovers}
\authoraddr{Corresponding author: Juan J. Valencia-Cardona,
Scientific Computing Program, University of Minnesota, Minneapolis, Minnesota, USA.
(valen167@umn.edu)}

\begin{document}

\title{Bullen's parameter as a seismic observable for spin crossovers in the lower mantle}

\authors{Juan J. Valencia-Cardona \altaffilmark{1},  Quentin Williams \altaffilmark{2}, Gaurav Shukla \altaffilmark{3},  Renata M. Wentzcovitch \altaffilmark{4,5}}

\altaffiltext{1}{Scientific Computing Program, University of Minnesota, Minneapolis, Minnesota, USA}
\altaffiltext{2}{Department of Earth and Planetary Sciences, University of California Santa Cruz, Santa Cruz, California, USA}
\altaffiltext{3}{Department of Earth, Ocean, and Atmospheric Science, Florida State University, Tallahassee, Florida, USA}
\altaffiltext{4}{Department of Applied Physics and Applied Mathematics, Columbia University, New York City, New York, USA}
\altaffiltext{5}{Lamont-Doherty Earth Observatory, Columbia University, Palisades, New York, USA}

\date{\today}

\clearpage

\begin{abstract}
Elastic anomalies produced by the spin crossover in ferropericlase have been documented by both first principles calculations and high pressure-temperature experiments. The predicted signature of this spin crossover in the lower mantle is, however, subtle and difficult to geophysically observe within the mantle. Indeed, global seismic anomalies associated with spin transitions have not yet been recognized in seismologic studies of the deep mantle. A sensitive seismic parameter is needed to determine the presence and amplitude of such a spin crossover signature. The effects of spin crossovers on Bullen's parameter, $\eta$, are assessed here for a range of compositions, thermal profiles, and lateral variations in temperature within the lower mantle. Velocity anomalies associated with the spin crossover in ferropericlase span a depth range near 1,000 km for typical mantle temperatures. Positive excursions of Bullen's parameter with a maximum amplitude of $\sim$ 0.03 are calculated to be present over a broad depth range within the mid-to-deep lower mantle: these are largest for peridotitic and harzburgitic compositions. These excursions are highest in amplitude for model lower mantles with large lateral thermal variations, and with cold downwellings having longer lateral length-scales relative to hot upwellings. We conclude that predicted deviations in Bullen's parameter due to the spin crossover in ferropericlase for geophysically relevant compositions may be sufficiently large to resolve in accurate seismic inversions of this parameter, and could shed light on both the lateral variations in temperature at depth within the lower mantle, and the amount of ferropericlase at depth.

\end{abstract}

\begin{article}

\section{Introduction} \label{sec:intro}

The adiabatic nature of the convecting mantle is a frequently used concept in the geophysical sciences. For instance, equation of state parameters, which are used to calculate the elastic and thermodynamic properties of minerals at mantle conditions, are commonly assumed to be adiabatic within the convecting mantle, e.g., the adiabatic bulk modulus and its derivative. However, various geodynamic simulations and seismological models \citep{Bunge, Dziewonski, ak135, Mattern,Matyska,Matyska2} suggest that the mantle is regionally nonadiabatic, particularly in the shallow and deep mantle regions, and in some cases, at mid lower mantle pressures. The latter is important because deviations from adiabaticity within the mantle provide insights into temperature gradients, heat flux, thermal history, thermal boundary layers, phase transitions, chemical stratification, and compositional heterogeneities. Therefore, knowledge about the degree of adiabaticity of the mantle helps us to constrain its composition and thermal structures related to mantle convection \citep{Matyska}. 

A common observable that quantifies the adiabaticity level of the mantle is Bullen's parameter, $\eta$. Introduced and developed by \cite{Bullen1, Bullen2}, $\eta$ is a measure of the ratio between the actual density increase with pressure within the Earth (as constrained by a combination of seismology, the Earth's moment of inertia, and mass) with respect to the profile derived from adiabatic self-compression. As such, it is expected to be unity where the mantle is homogeneous, adiabatic, and free of phase transitions. Thus, deviations of $\eta$ from unity (generally $\sim \pm$ 0.1 or less), indicate super(sub)adiabatic regions, and consequently, the presence of thermal boundary layers, compositional variations or phase transitions. Moreover, there is also the possibility that due to internal heating within the mantle, the mantle may be systematically subadiabatic \citep{Bunge}.

Evaluations of $\eta$ in geodynamic simulations are generally done by probing the parameter space associated with plausible convection models. This includes examining the effects of possible variations of the thermal conductivity, thermal expansion coefficient, viscosity, internal heating, and heat flux from the core, each of which directly impact the inferred geotherms \citep{Bunge, Matyska,Matyska2}. For instance, if internal heating is relatively significant, subadiabaticity is expected. Additionally, differences in elastic properties between individual phases within an aggregate can also produce variations in Bullen's parameter, and hence apparent deviations from adiabaticity. This is a bulk attenuation effect. Specifically, bulk attenuation phenomena are attributed to internal shear stresses generated from the local
mismatch of the elastic moduli of neighboring grains in a given aggregate. One formulation of bulk attenuation by \cite{Heinz} characterizes it through the ratio of the adiabatic bulk modulus K$_S$ and an effective modulus (Reuss bound) K$_E$, since the mantle can be assumed to be under hydrostatic pressure. Attenuation is a complicated problem to tackle, because it involves calculating complex moduli with an associated time dependency \citep{Heinz, Heinz1, Budiansky}. Such bulk attenuation effects are beyond the scope of this study, since the calculations we conduct are not time-dependent, but certainly needs to be addressed to understand systematic deviations of Bullen's parameter from 1. Here, we study how anomalies in bulk modulus  induced by spin crossovers affect the Bullen's parameter, and hence inferred adiabaticity of the lower mantle.

Elastic anomalies produced by the spin crossover in ferropericlase (fp) and bridgmanite (bdg), have been documented by both first principles calculations and high pressure-temperature experiments \citep{Badro2003,Speziale,Tsuchiya06, Wentzcovitch09, Wu4,Crowhurst,Marquardt,Antonangeli,Murakami2,Wu1,Wu2, Hsu11, Hsu12, Shukla2, Shukla16b}. The predicted signatures of this spin crossover in the lower mantle are subtle. Despite the fact that thermally induced velocity heterogeneities associated with this spin crossover appear to correlate statistically with seismic tomographic patterns observed in
deeply rooted plumes \citep{Wu2}, spherically averaged anomalies have not yet been
recognized in seismologic studies of the deep mantle. This may be due to difficulties associated with resolving gradual changes in the slopes of seismic velocities as a function of depth, and the trade-offs involved in seismic inversions of depth-dependent
velocity and density structures. In particular, velocity anomalies associated with the spin
crossover in fp are anticipated to span a depth range greater than 1000 km at mantle
temperatures. Thus, a sensitive seismic parameter is needed to determine the presence or
absence of this spin crossover signature, which would in turn shed light on the amount of ferropericlase
in the lower mantle. Bullen's parameter $\eta$ is an ideal candidate as it relates seismic wave
speeds with density variations, and sensitively records deviations from adiabaticity. Moreover, deviations from Bullen's parameter can be readily identified because it has a clear reference value (unity) in an adiabatic mantle that is heated from below. We calculated one dimensional perturbations of $\eta$ due to changes in composition, temperature, and spin crossover.  We achieved this by computing $\eta$ of different relevant mantle aggregates along their own adiabats. We also approximate lateral variations in temperature by modeling differing areas and temperature differences between upwellings and downwellings. The mantle phases of the aggregates considered are bridgmanite (bdg: Al- Fe- bearing MgSiO$_3$ perovskite), CaSiO$_3$ perovskite (CaPv), and ferropericlase (fp: (Mg,Fe)O). The aggregates have Mg/Si ratios that range from 0.82 to 1.56 and are harzburgite (Mg/Si $\sim$ 1.56) \citep{baker}, chondrite (Mg/Si $\sim$ 1.07) \citep{Hart}, pyrolite (Mg/Si $\sim$ 1.24) \citep{McDonough}, peridotite (Mg/Si $\sim$ 1.30) \citep{hirose}, and perovskite only (Mg/Si $\sim$ 0.82) \citep{Williams}. The predicted deviations in $\eta$ due to the spin crossover are comparable to previously reported variations \citep{Bunge,Matyska,Matyska2,Mattern}, and may be sufficiently large to turn up in accurate seismic inversions of this parameter.

\section{Method and Calculation Details} \label{sec:method}

We used bdg Mg$_{1-x}$Fe$^{2+}_{x}$SiO$_3$, (Mg$_{1-x}$Al$_{x}$)(Si$_{1-x}$Al$_{x}$)O$_3$, (Mg$_{1-x}$Fe$^{3+}_{x}$)(Si$_{1-x}$Al$_{x}$)O$_3$, (Mg$_{1-x}$Fe$^{3+}_{x}$)(Si$_{1-x}$Fe$^{3+}_{x}$)O$_3$  ($x=0$ and $0.125$) and fp Mg$_{1-y}$Fe$_{y}$O ($y = 0$ and $0.1875$) thermoelastic properties from \cite{Shukla15a, Shukla2} and \cite{Wu1}. Results for other $x$ and $y$ values were obtained by linear interpolation. All compositions account for the spin crossover in fp unless otherwise noted, i.e., bdg's iron (ferrous and/or ferric) is in the high spin (HS) state and fp is in a mixed spin (MS) state of HS and low spin (LS) states. For CaPv, we used thermoelastic properties from \cite{Kawai, Tsuchiya}, which were reproduced within the Mie-Debye-Gr\"uneisen  \citep{Stixrude} formalism. The mantle aggregates in this study, namely, harzburgite  \citep{baker}, chondrite \citep{Hart}, pyrolite  \citep{McDonough}, peridotite \citep{hirose}, and perovskititic only \citep{Williams}, are mixtures within the SiO$_2$ - MgO - CaO - FeO - Al$_2$O$_3$ system (ignoring alkalis and TiO$_2$ is not anticipated to resolvably affect the results). In addition, the Fe-Mg partition coefficient $K_D = \frac{x/(1-x-z)}{y/(1-y)}$ between bdg and fp, which is known to be affected by the spin crossover \citep{Irifune, piet}, was assumed to be  uniform throughout the mantle with a value of 0.5. Further details about these compositions can be found in \cite{valencia}.

The adiabats of the different minerals and aggregates were integrated from their adiabatic gradient, 

\begin{equation}  
\label{eq:agradient}
\left( \frac{\partial T}{\partial P} \right)_S =  \frac{\alpha V T}{ C_{p}}
\end{equation}

We denote the molar fraction, molar volume, molar mass, thermal expansion coefficient, and isobaric specific heat of the i$^{th}$ mineral in the mixture as $\mu_i$, $V_i$, $M_i$, $\alpha_i$, and $Cp_i$ respectively. The aggregate properties such as volume, thermal expansion coefficient, and isobaric specific heat are then $V=\sum_i \mu_iV_i$, $\alpha= \sum_i \alpha_i \mu_i V_i / V $, and $C_{p} = \sum_i \mu_i C_{p_i}$. The adiabatic aggregate bulk moduli $K_S$ were obtained from the Voigt-Reuss-Hill (VRH) average. Moreover, the aggregate density $\rho = \sum_i \mu_i M_i/V$ and seismic parameter $\phi = K_S/\rho $ were calculated along the aggregate adiabat, in order to compute adiabatic changes of density with respect to pressure as,

\begin{equation}  
\label{eq:bullen}
\eta =  \phi  \frac{d\rho}{dP}
\end{equation}

where $\eta$ is the Bullen's parameter. If $\eta =1$ the mantle is homogeneous and adiabatic, whereas values of $\eta > 1$ can indicate a phase change as $\rho$ varies more rapidly with depth than predicted by the adiabat. Furthermore, values of $\eta < 1$ may signify the presence of a thermal boundary layer or substantial internal heat production. Details about equation (\ref{eq:bullen}) can be found in the supplementary information.

\newpage

\section{Results and Discussion} \label{sec:result}

\subsection{Observations of $\eta$ in the lower mantle}

\begin{figure}
\centering
\resizebox{16cm}{!}{\includegraphics{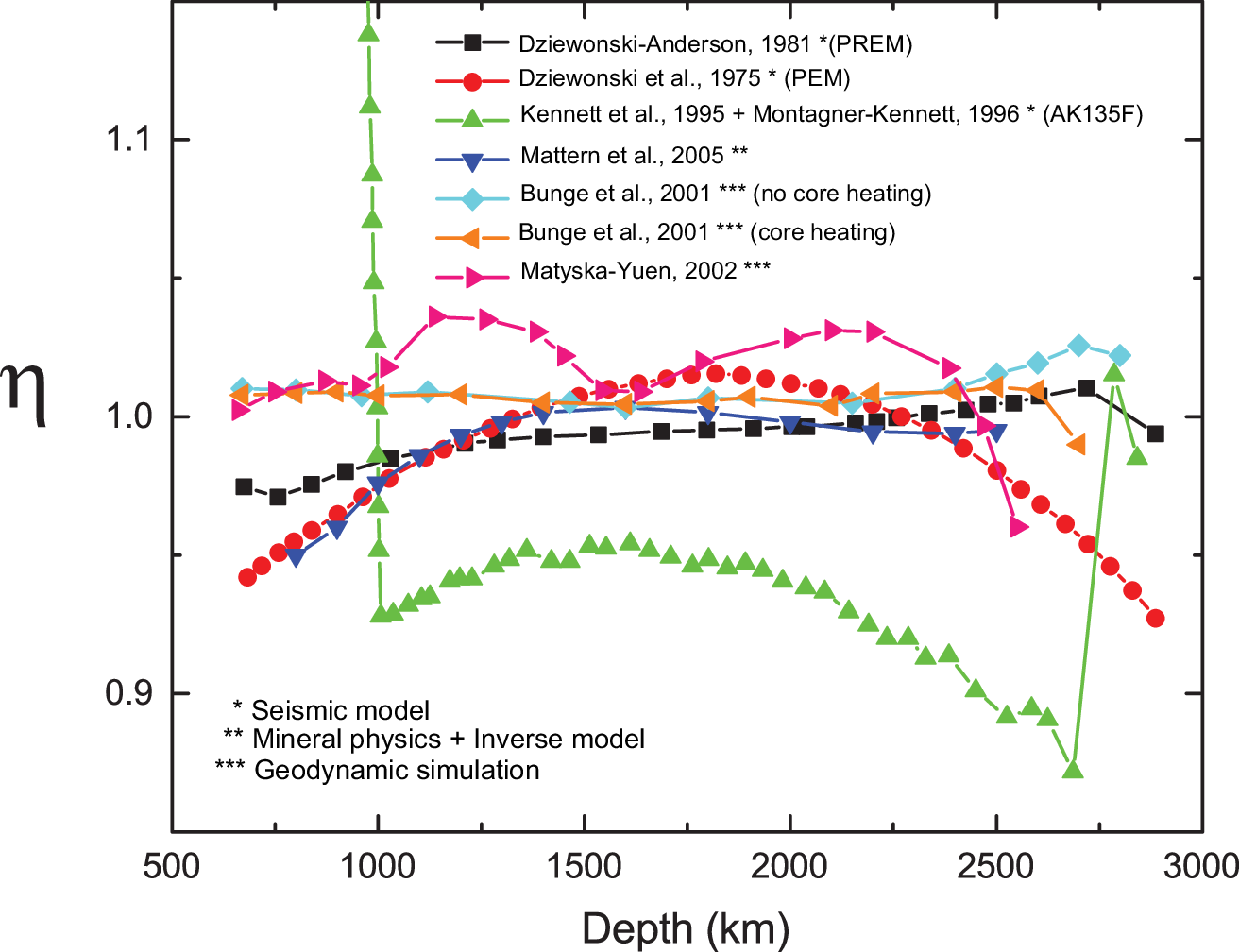}}
\caption{ Bullen's parameter $\eta$ calculations for seismic, geodynamic, and mineral physics models. }
\label{fig:models}
\end{figure}

Figure \ref{fig:models} shows different $\eta$ calculations from previous geodynamic \citep{Bunge,Matyska,Matyska2}, seismic \citep{dpem,Dziewonski,ak135}, and seismic plus mineral physics models with a priori starting conditions \citep{Mattern}.  Overall, $\eta$ oscillates between values of $\sim$ 1.04 to 0.96 for most of these models, except for the AK135F model \citep{ak135,Montagner}, which displays the largest fluctuations. AK135F exhibited an average value of $\eta$ $\sim$ 0.92 from 1000 km to 2700 km in depth,  and variations in $\eta$ above and below those depths were at least of the order of $\sim$ 0.1, which is substantially larger than the other inversions and calculations. For other seismic models such as PEM \citep{dpem} and PREM \citep{Dziewonski} such large fluctuations are not observed, but they could be suppressed by the continuity requirements of the polynomial formulations of these models. However, the Bullen's parameters of these seismic models do suggest the presence of a thermal boundary layer at the bottom of the lower mantle, as shown by the negative slope of all models in the bottommost hundred to few hundred km of the mantle. Notably, for the mineral physics plus inverse model calculation by \cite{Mattern}, $\eta$ values less than one from 800 km to 1300 km were attributed to iron depletion from their initially pyrolitic compositional model. 

Two and three dimensional geodynamic calculations of $\eta$ were first done by \cite{Matyska,Matyska2}, where the effect of varying parameter space properties, such as thermal conductivity, thermal expansion coefficient, and viscosity, lead to different perturbations in $\eta$, but with an average value of $\sim$ 1.01. This average value is in general agreement with other geodynamic calculations by \cite{Bunge}, which also showed that the presence of internal heat sources lead to subadiabatic regions. Other thermal contributions, like core heating, cause superadiabatic temperature gradients at the bottom of the mantle and thus the presence of a thermal boundary layer, as manifested by the negative slopes of $\eta$ near the base of the mantle. 

\subsection{Spin-crossover effect on the adiabaticity of the lower mantle}

\begin{figure}
\centering
\resizebox{16cm}{!}{\includegraphics{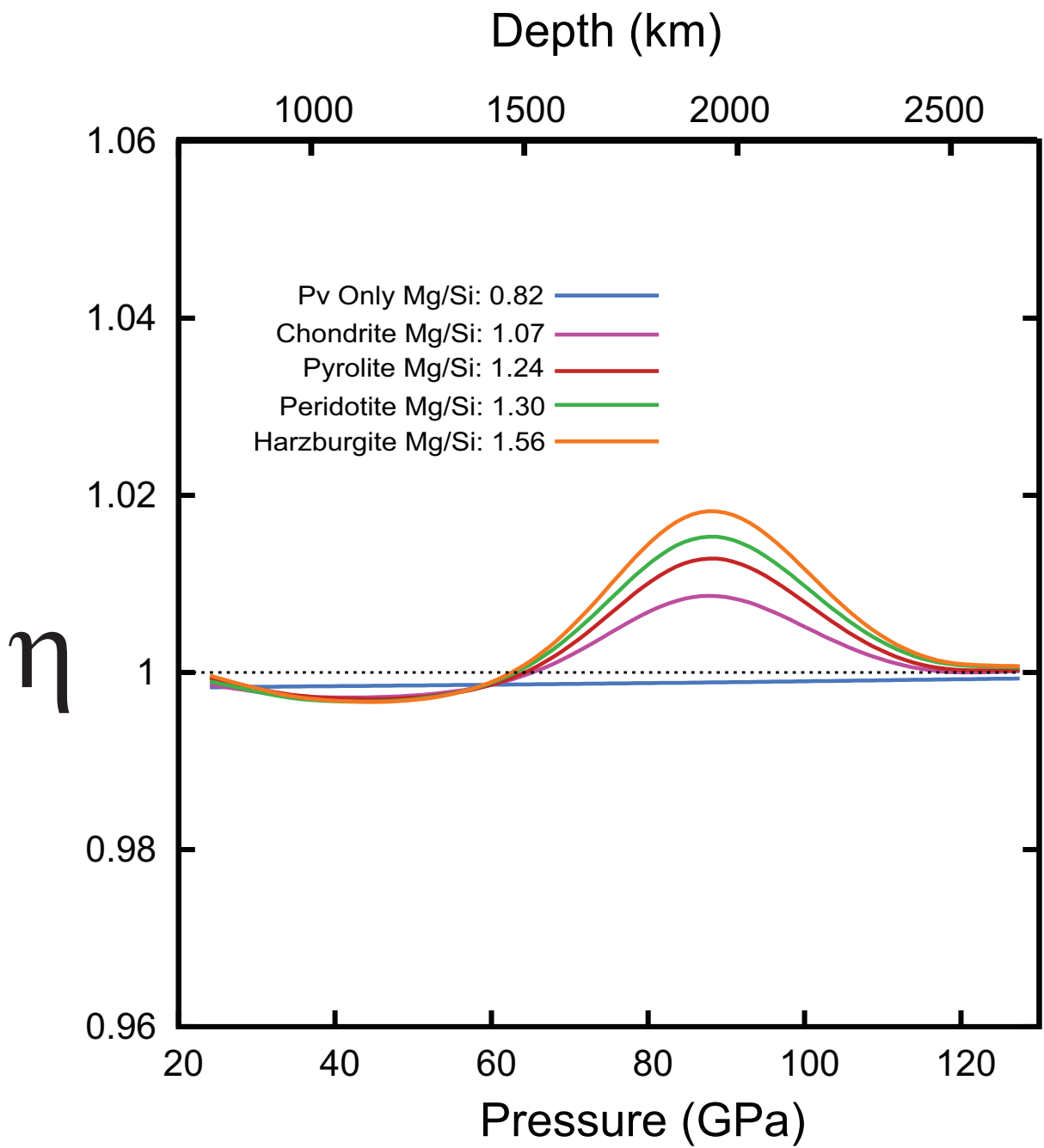}}
\caption{Perturbations of $\eta$ due spin crossover in fp in lower mantle aggregates. }
\label{fig:bullen_compositions}
\end{figure}

 We studied the effect of spin crossovers on lower mantle adiabaticity by examining $\eta$ excursions for different lower mantle aggregates along their self consistent adiabats. All of the adiabats of the different aggregates are listed in \cite{valencia}. Figure \ref{fig:bullen_compositions} shows the variations of $\eta$ only due to the spin crossover in fp: only a portion of trivalent iron in bdg is anticipated to undergo a spin transition within the mantle (e.g., \cite{catalli} and \cite{Hsu11, Hsu12}). For compositions with fp, fluctuations in  $\eta$ were $\sim$ 0.02 max, which are well within the variations in seismological observations and geodynamical calulations shown in Figure \ref{fig:models}. Furthermore, larger deviations from adiabaticity occur as the aggregate's Mg/Si ratio, i.e., fp content, is increased. The sensitivity to Mg/Si content of the Bullen's parameter maximum near 1900 km depth, induced by the spin crossover, is relatively large: peridotitic and harzburgitic compositions have an  $\eta$ anomaly which is nearly twice that of the chondritic composition. The $\eta$ excursions for the perovskitic composition, Pv only, depict the profile of a composition without fp in the lower mantle.

\subsection{Lateral temperature variations}

\begin{figure}
\centering
\resizebox{18cm}{!}{\includegraphics{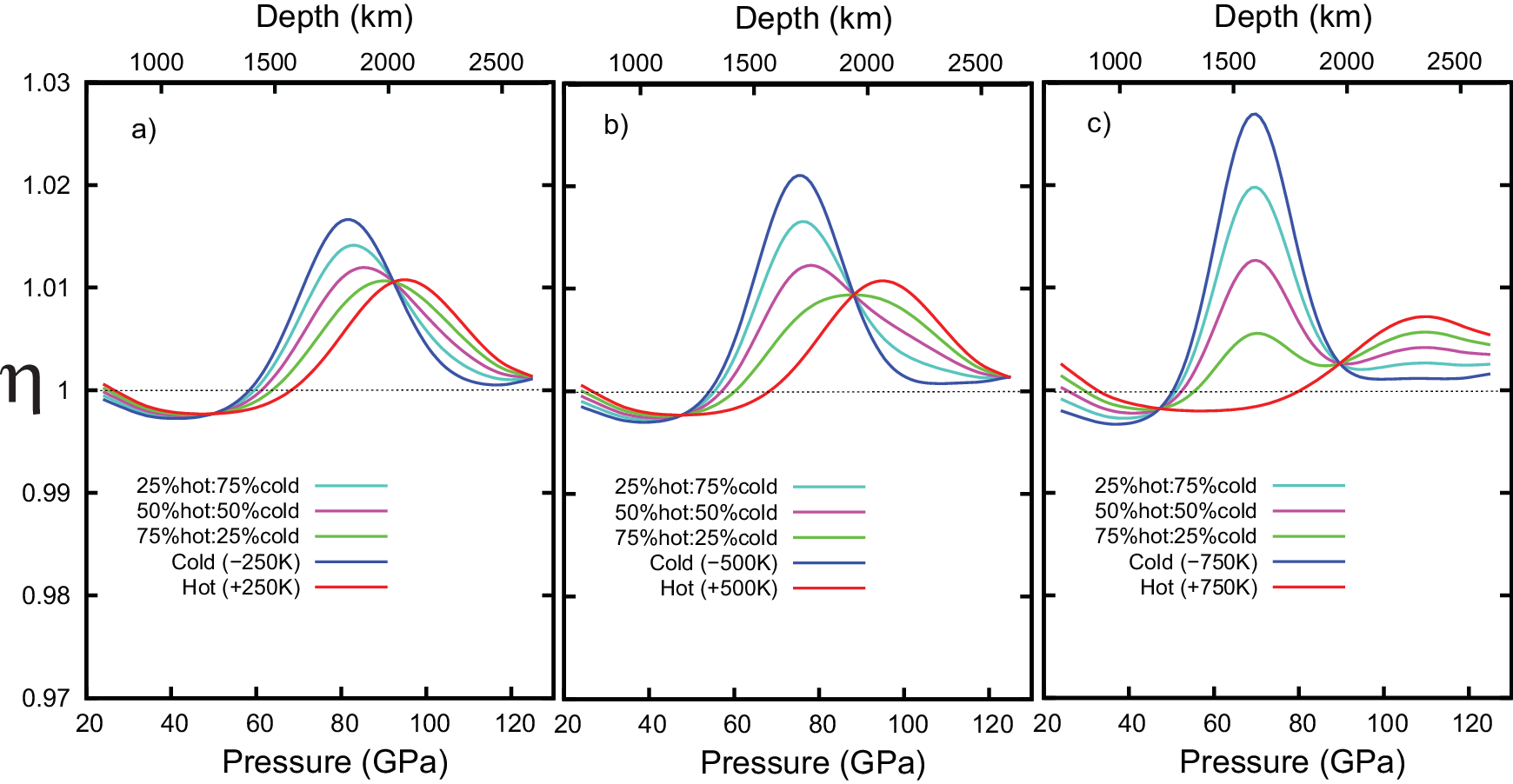}}
\caption{Lateral temperature variations of a) $\pm$ 250 K, b) $\pm$ 500 K, and c) $\pm$ 750 K for sequences of 25\%:75\%, 50\%:50\% and 75\%:25\% of the mantle being hot:cold.} In accord with the two-state model for temperature that we have assumed, two isosbestic points are generated near 1250 km and 2000 km depth, respectively.
\label{fig:averages}
\end{figure}

We have characterized what Bullen parameter anomalies, due to spin crossovers, might generate for one-dimensional seismic models of an isochemical adiabatic mantle. However, the lack of maxima in most Bullen parameter observations (Figure \ref{fig:models}) that are at the appropriate depth and have the right breadth to correspond to the spin crossover of fp, led us to probe the effect of lateral temperature variations on deviations of  $\eta$. Since lateral temperature variations and their areal distribution at a given depth of hot/upwelling and cold/downwelling material are not well-constrained in the deep mantle (e.g., \cite{Houser}), we conducted a sensitivity analysis for the effect of thermal variations on $\eta$ in a pyrolitic mantle. Here, material at each depth is distributed along adiabats with potential temperatures above (hot) and below (cold) a reference
adiabat pinned at 1873 K at 23 GPa as in \cite{Brown} (B\&S) (See also \citep{Akaogi}). The lateral temperature variations between hot and cold regions that we probed were  $\pm$ 250 K, $\pm$ 500 K, and $\pm$ 750 K in a sequence of 25\%:75\%, 50\%:50\% and 75\%:25\% ratio of the mantle at a given depth being hot:cold (See Figure \ref{fig:averages}).

For all the temperature-average distributions (Figures \ref{fig:averages}a, \ref{fig:averages}b, and \ref{fig:averages}c), we observed that the spin crossover anomalies, i.e. deviations from adiabaticity, became more prominent at lower temperatures: this is a natural consequence of the broadening of the spin transition that occurs at high temperatures. Conversely, greater amounts of hot material tend to make spin crossovers more difficult to resolve. Furthermore, we also observed that for large temperature variations, $\pm$ 750 K, two peaks in $\eta$ can also be generated at different depths in an isochemical thermally heterogeneous mantle (Figure \ref{fig:averages}c). This phenomenon is attributed to the volume increase with temperature, which increases the pressures that are required for the spin crossover to occur. Since the amplitude of the perturbations in $\eta$ increases also with higher fp content, it is expected that regions with larger cold harzburgitic chemistry present within the lower mantle, such as subducting slabs, should have substantially greater local fluctuations in the Bullen's parameter if a local vertical sampling of $\eta$ over such regions is performed. 

\begin{figure}
\centering
\resizebox{14cm}{!}{\includegraphics{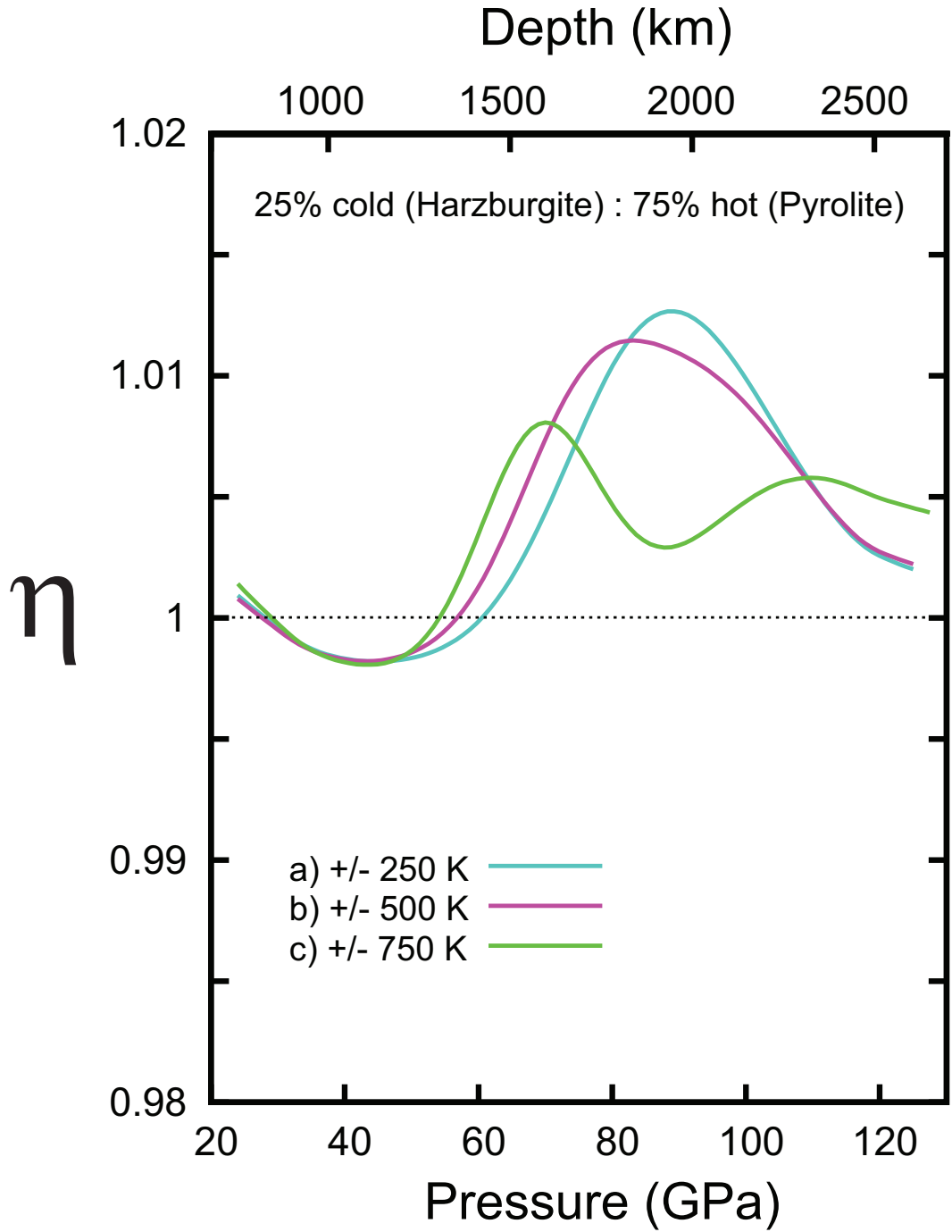}}
\caption{Lateral temperature and composition variations of a) $\pm$ 250 K, b) $\pm$ 500 K, and c) $\pm$ 750 K for a mantle being  25\% cold(harzburgite): 75 \% hot(pyrolite).}
\label{fig:averages_harz_pyro}
\end{figure}

Beyond lateral temperature variations, we examined the case of coupled compositional and thermal lateral heterogeneities. The rationale here is that cold, downwelling subducted material is likely to have a larger concentration of harzburgite than ambient mantle. We utilized a similar temperature averaging scheme, but with cold $\eta$ values being harzburgitic. Figure \ref{fig:averages_harz_pyro}  
shows different $\eta$ profiles with the mantle being 75\% hot(pyrolite) and 25\% cold(harzburgite). For this scenario, perturbations in $\eta$ due to the spin crossover vary their magnitude and reach a maxima at different depths, depending on the temperature difference between the cold downwellings of harzburgitic chemistry and ambient pyrolitic mantle. If the temperature difference is sufficiently large, e.g. $\pm$ 750 K, multiple peaks can be observed. Thus, the relative amplitudes and locations of multiple peaks could, if observed/observable, provide strong constraints on lateral variations in the geotherm and/or composition of the deep mantle. In particular, the depth at which the spin transition-induced peak occurs in Bullen's parameter is highly sensitive to temperature (Figure \ref{fig:averages}), while the amplitude of its variation is sensitive to composition (Figure \ref{fig:bullen_compositions}).

\section{Geophysical Significance} \label {Geophys}

We have utilized $\eta$ as an observable for spin crossovers in the lower mantle for the first time, in an attempt to reconcile mineral physics with seismic observations and to understand how such spin crossovers may affect observations of deviations from adiabaticity within the mantle. Our results suggest that the spin crossover signatures in $\eta$ should be sufficiently large to turn up in accurate (ca. 1\%) seismic inversions for this parameter. Whether such accuracies are achievable is unclear: several decades ago, \cite{Masters1979} concluded that $\eta$ variations from seismic observations could be resolved with a precision no better than 2\%.
Recent results from an inverse Bayesian method, deployed via a neural network technique by \cite{dewit}, showed that $\rho$, Vp, and Vs may each be resolvable to somewhat better than 1\% in the $\sim$ 2000 km depth range, based on their observed probability density functions. A linear combination of these uncertainties will certainly lead to values of order 1-2.5\% for the net uncertainty in 1-D inversions for Bullen's parameter. Nevertheless, given markedly improved and more accurate seismic inversions coupled with substantially larger data sets, it is possible that better constraints on $\eta$ might be developed.  

We also highlight the importance of the chosen temperature profile, as it has a direct impact on $\eta$. Elastic moduli, seismic velocities, and aggregate densities strongly  depend on  temperature. Hence, super(sub)adiabatic geotherms will lead to different interpretations of $\eta$. As recently showed by \cite{valencia},  the spin crossover in fp and bdg induces an increment in the adiabat's temperature of a given aggregate and such a temperature increment will impact  $\eta$'s sensitivity. Because of the potentially complex coupling of lateral temperature differences with compositional variations, further work on the effect of spin crossovers on $\eta$ would likely benefit from an assessment within a three dimensional convective scheme, such as the formulation proposed by \cite{Matyska2}.

\section{Conclusions} \label{sec:conc}

Apparent deviations from adiabaticity due to spin crossover, as recorded by the Bullen's parameter, increased in proportion to the aggregate's ferropericlase content. The magnitude of these perturbations is generally consistent with the magnitude of variations in $\eta$ present in previous seismological and geodynamic inversions of $\eta$ in the lower mantle. Our results provide a sense of how much of a perturbation in $\eta$, given the spin crossover and lateral temperature variations, might be expected in one dimensional seismic models, with the net result being of order 1-2\%. Accurate characterization of $\eta$ either globally or locally could provide constraints on both the lateral temperature distribution and the fp content at depth, although such determinations hinge critically on achieving sufficient seismic resolution to resolve spin transitions. Also, the perturbations found in $\eta$ for different mantle temperature averages highlight the importance of doing vertical seismic velocity profiles with sufficient precision to allow $\eta$ to be characterized on a regional basis. Our results provide a guide for possible a priori models of $\eta$ in regionalized inversions of velocity as a function of depth: inversions without spin crossover induced perturbations in $\eta$ implicitly assume that spin transitions are absent at depth, and hence that no ferropericlase is present in the deep mantle.

\begin{acknowledgement}
We thank two reviewers for helpful comments. This work was supported primarily by grants NSF/EAR 1319368 and 1348066. Q. Williams was supported by NSF/EAR 1620423. Results produced in this study are available in the supporting information. The 2016 CIDER-II program (supported by NSF/EAR 1135452) is thanked for providing a portion of the original impetus of this study.
\end{acknowledgement}

\section{Supplementary Material}

The supporting information consists of Text S1, Figure S1, and Table S1. 
Figure S1 shows $\eta$ for prystine lower mantle minerals, namely, MgSiO$_3$, MgO, and CaSiO$_3$. 
Text S1 shows the Bullen's parameter $\eta$ derivation. 
Table S1 the values of $\eta$ for the different aggregates. 

\subsection{Text S1}

The bulk modulus of a mineral under adiabatic self compression is given by,

\begin{equation}
 K_S = \rho \left( \frac{\partial P}{\partial \rho} \right)_S
\label{eq:s1}
 \end{equation}

Hence, 

\begin{equation}
 \frac{K_S}{\rho} = \left( \frac{\partial P}{\partial \rho} \right)_S = \phi
 \label{eq:s2}
\end{equation}

where $\phi$ is the seismic parameter $\phi = V_P^2 - \left( \frac{4}{3}\right)V_S^2 $.

Furthermore, assuming a homogeneous media under hydrostatic changes in pressures with respect to depth, 

\begin{equation}
 \frac{dP}{dr} = -\rho g 
\label{eq:s3}
 \end{equation}

where $g$ and $r$ are the acceleration due gravity and depth, respectively. 

Thus, 

\begin{equation}
 \frac{dP}{d\rho} \frac{d\rho}{dr} = -\rho g 
\end{equation}

and using equation (\ref{eq:s2}),
\begin{equation}
 1 = - \phi \rho^{-1} g^{-1} \frac{d\rho}{dr}
 \label{eq:s4}
\end{equation}

Equation (\ref{eq:s4}) is known as the Adams-Williamson equation. 

If the system is not adiabatic, i.e., equation (\ref{eq:s4}) differs from the unity, we have,

\begin{equation}
 \eta = - \phi \rho^{-1} g^{-1} \frac{d\rho}{dr} = \phi \frac{d\rho}{dP}
 \label{eq:s5}
\end{equation}

where $\eta$ is the Bullen's parameter \citep{Bullen2}.

\begin{figure}
\setfigurenum{S1} %%Change number for each figure
\centering
\noindent\includegraphics[width=17cm]{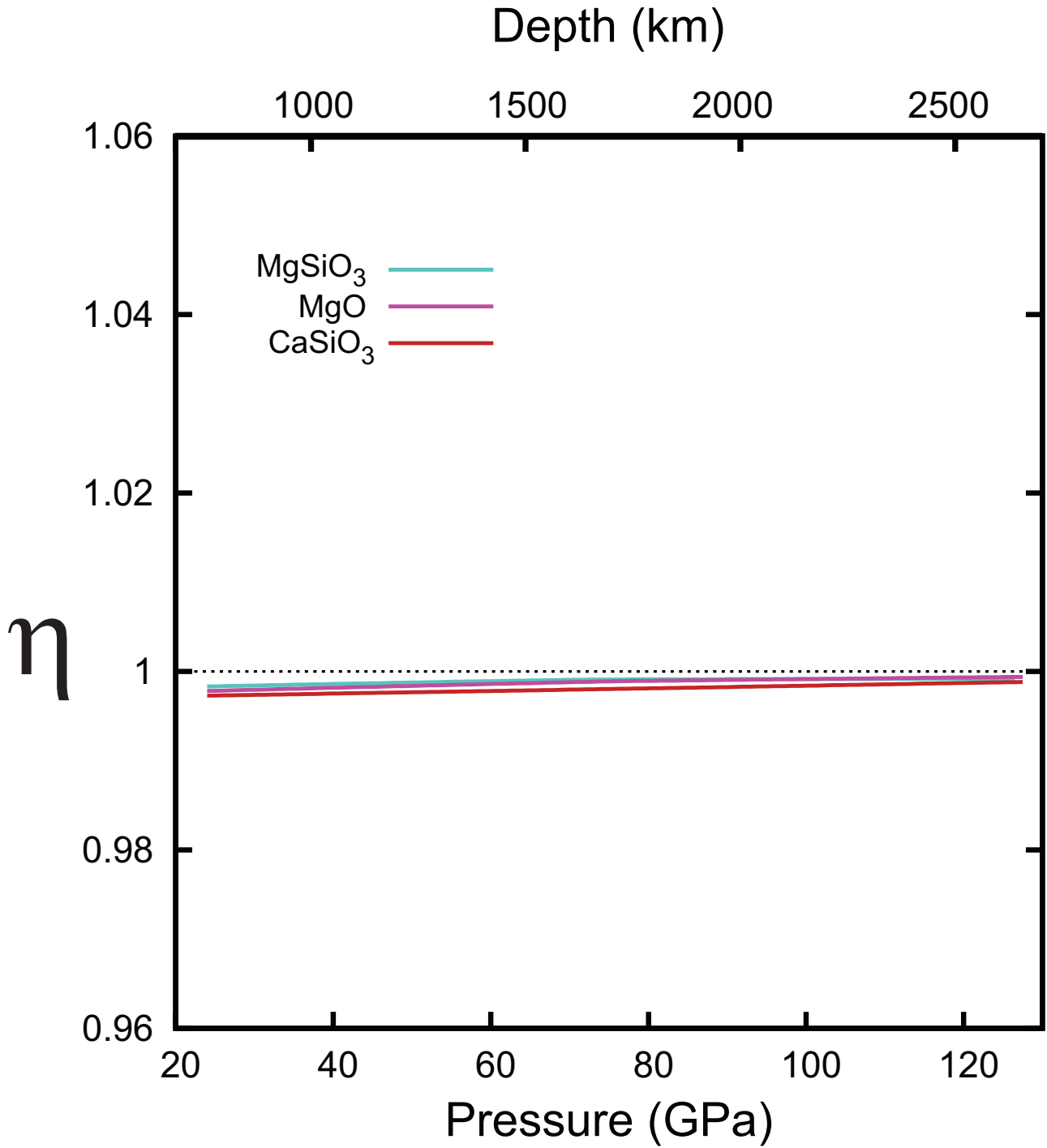}
	\caption{Bullen's parameter $\eta$ for pristine lower mantle minerals.}
	\label{figure1}
\end{figure}

\end{article}

\begin{landscape}

\begin{table}
\caption{Bullen's parameter $\eta$ for aggregates with fp in MS state.}
\label{tbl:bullen}
\resizebox{22cm}{!} {
\begin{tabular}{lccccccccccccccc}\hline \\
				&    &   Perovskite Only    &	 &	& Chondritic   &           &    &  Pyrolite  &  &   &        Peridotite   &&&      Harzburgite    \\ 
$\qquad$   Pressure  (GPa)    	&     &	 $\eta$ &    &	  &  $\eta$ & &	&  $\eta$ &  &  &   $\eta$         &&&     $\eta$ &        \\ \hline \\
$\qquad$ $\qquad$ 23		&    &  0.9983      &    $\qquad$     &       &  0.9989    &	 $\qquad$  &     &  1.0008        &   $\qquad$   &     & 0.9990    &  $\qquad$     &       & 0.9996   &      \\
$\qquad$ $\qquad$ 30	        &    &  0.9983      &    $\qquad$     &       &  0.9981    &	 $\qquad$  &     &  0.9996        &   $\qquad$   &     & 0.9978    &  $\qquad$     &       & 0.9981   &      \\
$\qquad$ $\qquad$ 40	        &    &  0.9984      &    $\qquad$     &       &  0.9974    &	 $\qquad$  &     &  0.9985        &   $\qquad$   &     & 0.9967    &  $\qquad$     &       & 0.9967   &      \\
$\qquad$ $\qquad$ 50	        &    &  0.9985      &    $\qquad$     &       &  0.9976    &	 $\qquad$  &     &  0.9985        &   $\qquad$   &     & 0.9968    &  $\qquad$     &       & 0.9968   &      \\
$\qquad$ $\qquad$ 60	        &    &  0.9986      &    $\qquad$     &       &  0.9987    &	 $\qquad$  &     &  0.9998        &   $\qquad$   &     & 0.9986    &  $\qquad$     &       & 0.9987   &      \\
$\qquad$ $\qquad$ 70	        &    &  0.9987      &    $\qquad$     &       &  1.0020    &	 $\qquad$  &     &  1.0041        &   $\qquad$   &     & 1.0038    &  $\qquad$     &       & 1.0046   &      \\
$\qquad$ $\qquad$ 80	        &    &  0.9987      &    $\qquad$     &       &  1.0073    &	 $\qquad$  &     &  1.0116        &   $\qquad$   &     & 1.0125    &  $\qquad$     &       & 1.0148   &      \\
$\qquad$ $\qquad$ 90	        &    &  0.9990      &    $\qquad$     &       &  1.0091    &	 $\qquad$  &     &  1.0145        &   $\qquad$   &     & 1.0157    &  $\qquad$     &       & 1.0187   &      \\
$\qquad$ $\qquad$ 100	        &    &  0.9990      &    $\qquad$     &       &  1.0054    &	 $\qquad$  &     &  1.0092        &   $\qquad$   &     & 1.0097    &  $\qquad$     &       & 1.0117   &      \\
$\qquad$ $\qquad$ 110	        &    &  0.9991      &    $\qquad$     &       &  1.0016    &	 $\qquad$  &     &  1.0036        &   $\qquad$   &     & 1.0032    &  $\qquad$     &       & 1.0038   &      \\
$\qquad$ $\qquad$ 120	        &    &  0.9992      &    $\qquad$     &       &  1.0002    &	 $\qquad$  &     &  1.0016        &   $\qquad$   &     & 1.0007    &  $\qquad$     &       & 1.0009   &      \\
$\qquad$ $\qquad$ 125		&    &  0.9992      &    $\qquad$     &       &  1.0003    &	 $\qquad$  &     &  1.0016        &   $\qquad$   &     & 1.0006    &  $\qquad$     &       & 1.0007   &      \\ \hline\hline		
\end{tabular}                                                                                                                                                                            
}
\end{table}

\end{landscape}

\end{document}